\documentclass[twoside]{article}
\usepackage{fleqn,espcrc2}
\usepackage{here,graphicx}
\usepackage{latexsym}
\newcommand{\figscale}{0.51}

\title{Non-trivial phase structure of $N_f=3$ QCD with $O(a)$-improved Wilson 
fermion at zero temperature
\thanks{Presented by M. Okawa.}
}

\author{
  JLQCD Collaboration:
  S.~Aoki\address{Institute of Physics, University of Tsukuba, 
                  Tsukuba, Ibaraki 305-8571, Japan\\[-1ex]},
  R.~Burkhalter$^{\rm a,}$\address{Center for Computational Physics,
               University of Tsukuba, Tsukuba, Ibaraki 305-8577,
               Japan\\[-1ex]},
  M.~Fukugita\address{Institute for Cosmic Ray Research, 
                    University of Tokyo, Kashiwa, Chiba 277-8582, Japan\\[-1ex]},
  S.~Hashimoto\address{High Energy Accelerator Research Organization~(KEK),
                       Tsukuba, Ibaraki 305-0801, Japan\\[-1ex]},
  K-I.~Ishikawa$^{\rm d}$,
  N.~Ishizuka$^{\rm a,b}$,
  Y.~Iwasaki$^{\rm a,b}$,
  K.~Kanaya$^{\rm a,b}$,
  T.~Kaneko$^{\rm d}$,
  Y.~Kuramashi$^{\rm d}$,
  M.~Okawa$^{\rm d}$,
  T.~Onogi\address{Yukawa Institute for Theoretical Physics,
                   Kyoto University, Kyoto 606-8502, Japan\\[-1ex]},
  S.~Tominaga$^{\rm b}$,
  N.~Tsutsui$^{\rm d}$,
  A.~Ukawa$^{\rm a,b}$,
  N.~Yamada$^{\rm d}$,
  T.~Yoshi\'e$^{\rm a,b}$
  }

\begin{document}
\begin{abstract}

JLQCD collaboration recently started the $N_f=3$ QCD simulations with 
the $O(a)$-improved Wilson fermion action employing an exact fermion algorithm 
developed for odd number of quark flavors.  It is found that this theory 
has an unexpected non-trivial phase structure in the $(\beta,\kappa)$ plane 
even at zero temperature.  
A detailed study is made to understand the nature of the 
observed phase transitions and to find the way of avoiding untolerably large 
lattice artifacts associated with the phase transition.

\vspace{-120mm}
\begin{flushright}
\large KEK-CP-111
\end{flushright}
\vspace{103mm}

\end{abstract}

\maketitle


\section{Introduction}
Including the dynamical quark loop effects in large-scale QCD simulations 
is one of the most important and urgent problem.  While in the real world 
there are three light quarks, most of recent studies have concentrated 
on the case of two flavors of Wilson-type fermions 
because these systems can be simulated by the exact 
Hybrid Monte Carlo (HMC) algorithm.  
Recently, however, a major progress has been made in 
the exact fermion algorithm applicable to odd number of 
flavors as reviewed in Ref.~\cite{review}.  
In particular, we have succeeded in developing an efficient algorithm 
for $N_f=3$ QCD with the $O(a)$-improved Wilson fermion action
in the framework of the HMC algorithm.  
Leaving algorithmic details to a separate report~\cite{kishika}, 
we present here the first result toward realistic 
$N_f=3$ QCD simulations with the $O(a)$-improved Wilson fermion.
We found that this theory has an unexpected first-order phase transition 
at zero-temperature, which gives a strong constraint on the form of 
lattice actions suitable for large-scale simulations. 

\section{Phase structure for plaquette gauge action}

The partition function we study is defined by
\begin{equation}
{\cal Z}=\int\!\!{\cal D}U\,
                  (\det[D_{ud}])^2(\det[D_s])e^{-S_{g}(U)}.
\label{eq:pertition_function}
\end{equation}
Here $S_g(U)$ is the gluon action 
\begin{equation}
S_g(U)=
{\beta \over 6}\left[c_0 \sum W_{1 \times 1} + c_1\sum W_{1 \times 2}\right],
\label{eq:gluon_action}
\end{equation}
where $c_0=1-8c_1$ and $W_{1 \times 1}$ and $W_{1 \times 2}$ are the plaquette 
and rectangular Wilson loops, respectively, and    
$(\det[D_{ud}])^2$ represents the contribution from degenerate $u$ and $d$ 
quarks whereas $s$ quark effect is given by $(\det[D_s])$ with $D_q (q=ud,s)$ 
the $O(a)$-improved Wilson-Dirac operator.

We start our analysis employing the plaquette gauge 
action ($c_1=0$), determining the clover coefficient $c_{\mathrm{sw}}$ 
by tadpole-improved one-loop perturbation theory;
\begin{equation}
c_{\mathrm{sw}}={1 \over \langle P\rangle^{3 \over 4}} 
\left(1+0.0159{6/\beta \over \langle P\rangle} \right),
\label{eq:c_sw_plaqutte}
\end{equation}
with $\langle P\rangle$ evaluated in the quenched approximation 
at the corresponding value of $\beta$. 

Given the lattice action, we have three tunable 
parameters, the gauge coupling $\beta$, the hopping parameter 
$\kappa_{ud}$ for $ud$ 
quarks and $\kappa_s$ for $s$ quark.  For simplicity we set $\kappa_{ud}
=\kappa_s \equiv \kappa$ in this work.  
To see the global phase structure 
in the ($\beta,\kappa$) plane,  we perform rapid thermal cycles
on $4^3 \times 16$ and $8^3 \times 16$ lattices.  Fixing $\beta$  
we increase $\kappa$ from $\kappa=0$ in steps of 0.01 until 
the algorithm fails to converge and then decrease $\kappa$ to 0.  
The procedure is repeated for several values of $\beta$.  In this  
analysis, we omit the noisy Metropolis test for the correction 
factor $\det[W_{oo}]$~\cite{kishika} 
and fix the order of the Chebyshev polynomial to $n=100$.  

\begin{figure}
\vspace*{-1ex}
\centering
\hspace*{-1em}
\includegraphics[scale=\figscale,clip]{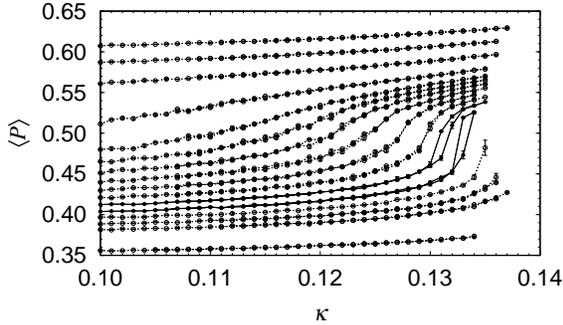}
\vspace{-9ex}
\caption{Thermal cycles on an $8^3 \times 16$ lattice at 
$\beta$ = 4.6, 4.8, 4.85, 4.9, 4.95, 5.0, 5.05, 
5.1, 5.15, 5.2, 5.25, 5.3, 5.4, 5.6, 5.8, 6.0 from bottom to top.}
\label{fig:fig1}
\vspace{-6ex}
\end{figure}
\begin{figure}
\vspace*{-1ex}
\centering
\hspace*{-1em}
\includegraphics[scale=\figscale,clip]{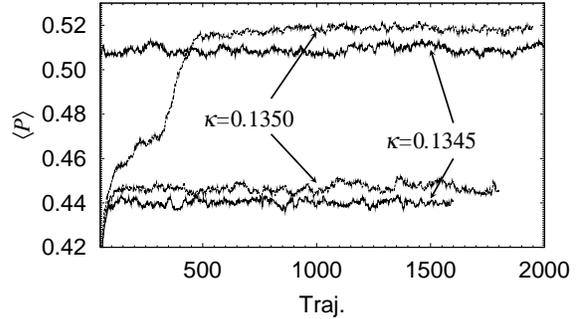}
\vspace{-9ex}
\caption{Two-state signals on a $12^3 \times 32$ lattice at $\beta$=4.88 
and $c_{\mathrm{sw}}$=2.15.}
\label{fig:fig2}
\vspace{-6ex}
\end{figure}

The result of the thermal cycle on an $8^3 \times 16$ lattice is shown in 
Fig.~\ref{fig:fig1}.  
We clearly observe hysteresis loops at $\beta=4.95$ and $5.0$.  
On a $4^3 \times 16$ lattice, the hysteresis loop is seen in a wider range of 
$\beta$ at $4.8 \le \beta \le 5.1$.  
To understand the nature of the hysteresis 
in more detail, we perform 
simulations starting from both ordered and disorder configurations with 
fixed values of $\beta$ and $\kappa$ on $8^3 \times 16$ and $12^3 \times 32$ 
lattices.  Some examples are shown in Fig.~\ref{fig:fig2}.  
Here we make simulations exact 
by including the correction factor $\det[W_{oo}]$.
We observe clear two state signals for several sets of parameters, 
which demonstrates that the hysteresis in the thermal cycles is 
caused by a first-order phase transition.  

\begin{figure}
\vspace*{-1ex}
\centering
\hspace*{-1em}
\includegraphics[scale=\figscale,clip]{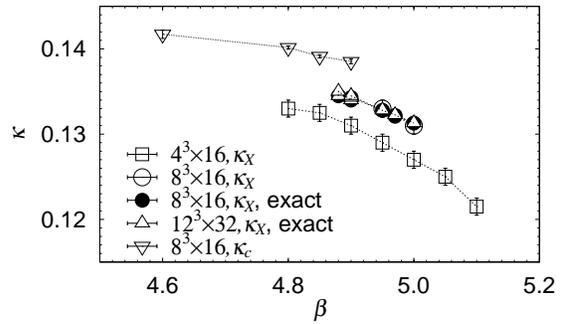}
\vspace{-9ex}
\caption{Phase diagram in $(\beta,\kappa)$ plane.}
\label{fig:fig3}
\vspace{-6ex}
\end{figure}
In Fig.~\ref{fig:fig3} we plot the location of the observed phase 
transition line ($\kappa=\kappa_X(\beta)$) 
on the ($\beta$, $\kappa$) plane for 
$8^3 \times 16$(filled circles) and $12^3 \times 32$ (open upper-triangles)
lattices. Also shown in Fig.~\ref{fig:fig3} are the location of the 
hysteresis observed on $4^3 \times 16$ and $8^3 \times 16$ lattices 
(open squares and circles).  
The position of the phase transition line significantly moves  
from $4^3 \times 16$ to $8^3 \times 16$.  
However, it stays in the same place when the lattice size  
increases from $8^3 \times 16$ to $12^3 \times 32$.
\begin{figure}[top]
\vspace*{-1ex}
\centering
\hspace*{-1em}
\includegraphics[scale=\figscale,clip]{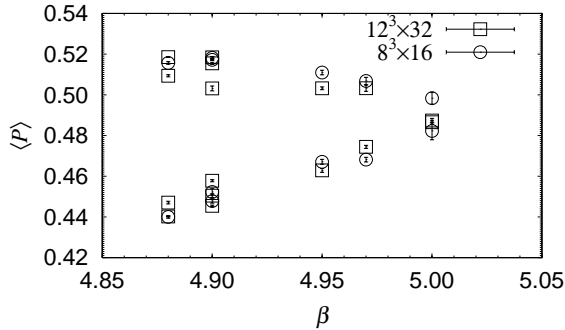}
\vspace{-9.5ex}
\caption{Plaquette values in two phases at the first-order transition line 
as functions of $\beta$.}
\label{fig:fig4}
\vspace{-6ex}
\end{figure}
In Fig.~\ref{fig:fig4}, we plot the plaquette values in the two phases 
at the phase transition line as functions of $\beta$ for 
$8^3 \times 16$ and $12^3 \times 32$ lattices.  The gap in the plaquette 
is almost independent of the lattice size.
These findings strongly suggests 
that the first-order transition line persists in infinite lattice 
volume and zero temperature.  

Figure~\ref{fig:fig4} also shows that the gap in the plaquette decreases 
almost linearly in $\beta$ and vanishes at around $\beta=5.0$.  
Since no signs of transitions are observed at $\beta > 5.0$ 
in Fig.~\ref{fig:fig1}, we do not 
think that this point corresponds to a tricritical point where a 
first-order phase transition line changes to a second-order one 
which then extends toward weaker couplings.  Rather, 
the observed first-order transition must be a lattice artifact 
restricted to strong coupling regions. 

\section{Phase structure for improved gauge action}
 
\vspace{-1ex}

To confirm this possibility, we repeat thermal cycle simulations 
employing the improved gauge action  
with $c_1=-0.331$ (RG action~\cite{iwasaki}) and 
$c_1=-1/(12\langle P\rangle^{1 \over 2})$ 
(tadpole-improved Symanzik action~\cite{symanzik}).  
Here $c_{\mathrm{sw}}$ is again determined by 
tadpole-improved one-loop perturbation theory.     
\begin{figure}
\vspace*{-1ex}
\centering
\hspace*{-1em}
\includegraphics[scale=\figscale,clip]{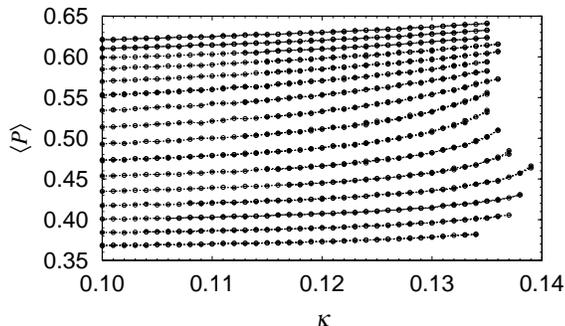}
\vspace{-9.5ex}
\caption{Thermal cycle with RG gauge action 
at $\beta$=1.50$-$2.25 in steps of 0.05 (bottom to top).}
\label{fig:fig5}
\vspace{-5ex}
\end{figure}
Figure~\ref{fig:fig5} shows the result of thermal cycles 
for the RG action on an 
$8^3 \times 16$ lattice.  No signs of hysteresis loops are seen, 
and hence non-trivial phase structure is absent with this gauge 
action. Hysteresis is also not seen for the tadpole-improved 
Symanzik gauge action.  
We conclude that the first-order phase transition observed for the 
plaquette gauge action has the nature of a lattice artifact.  

\vspace{-0.5ex}

\section{Implications}

\vspace{-0.5ex}

It has been known \cite{adjoint} that pure $\mathit{SU}(3)$ gauge theory 
with fundamental and adjoint couplings has a first-order transition 
for positive values of the adjoint coupling.  
It is an interesting possibility that the clover term 
produces a positive adjoint coupling in an effective theory after 
integrating out the fermion degrees of freedom, which eventually causes the 
first-order phase transition observed in this work.
In fact we have checked that there are no sign of phase transitions 
in thermal cycles made with the unimproved Wilson fermion action
 ($c_{\mathrm{sw}}=0$) and plaquette gauge action ($c_1=0$) on a 
$8^3 \times 16$ lattice, which supports our interpretation.

Our finding poses a serious practical problem 
whether realistic simulations with 
reasonable values of the lattice spacing $a$ are possible 
with the $O(a)$-improved Wilson fermion and the plaquette gauge action.  
To study this problem, we evaluate the 
lattice spacing from static quark potential at $\beta=5.0$.  Our results are 
$a^{-1}=1.53(2)$ GeV $(\kappa=0.1320)$, $2.06(5)$ GeV $(\kappa=0.1330)$ and 
$2.58(4)$ GeV $(\kappa=0.1338)$.  Continuum extrapolations 
using lattices with $a^{-1}$ starting at 2.58 GeV is not practical.
Our natural choice, then, is to use improved gauge actions for realistic 
$N_f=3$ QCD simulations with the $O(a)$-improved Wilson fermion.      

\vspace*{3mm}
This work is supported by the Supercomputer Project No. 66
(FY2001) of High Energy Accelerator Research Organization
(KEK), and also in part by the Grant-in-Aid of the Ministry
of Education (Nos. 10640246, 11640294,
12014202, 12640253, 12640279, 12740133, 13640260 and 13740169).
K-I.I. and N.Y. are supported by the JSPS Research
Fellowship.

\end{document}